# Non-Contextual and Local Hidden-Variable Model for the Peres-Mermin and Greenberger-Horne-Zeilinger Systems

Carsten Held[1]

(Dated: November 26, 2019)

A hidden-variable model for quantum-mechanical spin, as represented by the Pauli spin operators, is proposed for systems illustrating the well-known no-hidden-variables arguments by Peres and Mermin (1990) and by Greenberger, Horne, and Zeilinger (1989). Both arguments rely on an assumption of non-contextuality; the latter argument can also be phrased as a non-locality argument, using a locality assumption. The model suggested here is compatible with both assumptions. This is possible because the scalar values of spin observables are replaced by vectors that are components of orientations.

## I. INTRODUCTION

Before the creation of quantum mechanics (QM), measurement of a physical system was conceived as faithful in the sense that it consists in 'the ascertaining of some pre-existing property of some thing, any instrument involved playing a purely passive role' [1, 2]. As is now well-known, this classical conception of measurement is in conflict with QM, understood as a complete description of the quantum domain. Consider an interpretation of QM that maintains: all observables are faithfully measured in the sense that measurement reveals 'some pre-existing property' of the quantum system, 'any instrument involved playing a purely passive role'. This interpretation is in an immediate conflict with well-known no-hidden-variable arguments of the Bell-Kochen-Specker type [3, 4, 5, 6]. Such arguments show that any hidden-variable model for QM observables is necessarily contextual, given that algebraic relations among operators representing the observables are mirrored in the observables' values. Contextuality here means that the model must allow some observables to have different values in different contexts, i.e. as elements of different sets of observables.

What sense could be made of such contextuality? A natural idea would be measurement contextuality: the idea that the 'instrument' does *not* play 'a purely passive role' in the measurement process. An observable's value would thus depend upon the process of measurement of a set of observables including it. This idea, of course, conflicts with the original idea of measurement as the recording of pre-existing properties and thus, for the hidden-variable-theorist, is not worth pursuing. An alternative is ontological contextuality [7, 8], the idea that instead of one observable with different context-dependent values there are really two different observables (represented by the same operator) with different values. Without further explanation, this idea appears to be entirely ad hoc and, since no explanation has been forthcoming, is no longer pursued in the literature. It is widely agreed that the ideal of faithful measurement implies faithful measurement *of non-contextual properties* and as such is refuted by the mentioned arguments.

However, abandoning the idea of faithful measurement does not avoid contextuality. Since a QM system's being in a certain state is equivalent to the state's being an eigenstate of certain observables, being in a certain state is equivalent to being in a certain context of observables. If it is impossible that all observables have the same values in all states and some observables have values in certain states, others have no values in these states. So, whether an observable has a value depends upon which other observables have values. In this sense, QM is contextual even without faithful measurement. But we have no real understanding of this contextuality.

If there is contextuality it also exists among observables pertaining to space-like separated parts of a QM system [9, 10]. Non-contextuality of such observables is equivalent with their locality, and Bell's Theorem [11, 12] famously shows such locality to yield statistical predictions at variance with those of QM. Thus, contextuality reappears, in the context of Bell's Theorem, as non-locality and again it appears as a characteristic of QM that we do not really understand.

---

[1] E-mail: carsten.held@mailbox.org

Or is there no need here for further understanding, just one for acceptance? The majority of interpreters embraces a fully non-classical world-picture: non-faithful measurement, contextuality and non-locality. There is a minority of dissenters, interpreters trying to restore classicality, but they usually are forced to make extravagant metaphysical assumptions. Here are two examples from the recent literature. The consistent histories approach [13, 14, 15, 16] claims that it can save the mentioned three elements of classicality [17, 18] but it is forced to claim that there exists no single description of a QM system that is both exhaustive and true [19]. This approach, if indeed it saves classicality, does so at an exceedingly high price. We might think that classicality and the possibility of a single true and exhaustive description of any system are parts of the same realist world-view but according to this approach they are not. A second example comes from the group of no-detection theories, i.e. interpretations exploiting the so-called detection loophole. One of these theories, called extended semantic realism [20, 21, 22], explicitly claims to save the three classicality features but it must assume that the ensemble of detected systems always is an unfair sample from the one of prepared systems [23]. This suggests that Nature purposefully hides itself from the physicist. (Moreover, there is empirical evidence for the claim that the detection loophole can be closed [24], such that unfair sampling no longer plays an important role in the interpretation QM.)

In the following, a new attempt to save classicality will be made – but one that avoids counter-intuitive metaphysical consequences. A hidden variable will be suggested that effectively replaces QM observables and their scalar values by vector variables and their vector values. The result will be a conception of properties of QM systems respecting the three features of classicality. The conception will be developed for two related and particularly simple cases: the Peres-Mermin (PM) and Greenberger-Horne-Zeilinger (GHZ) systems consisting, respectively, of two and three spin-½ particles. In the resulting model, the QM observables are replaced by vector variables; these variables jointly have values that are identical across contexts, thus satisfying non-contextuality, and yet they meet the PM and GHZ constraints. Moreover, in the GHZ case the values of variables pertaining to an individual particle can be predicted without interfering with the particle, thus satisfying locality, and yet they meet the GHZ constraints on the values in a particular three-particle state. As a result, arguments from QM against faithful measurement no longer apply and we obtain a possibility to reclaim classicality.

## II. THE PM, GHZ AND BELL-GHZ ARGUMENTS

We consider no-hidden-variables arguments that employ systems consisting of two or three spin-½ particles described by the familiar Pauli spin operators. We first recall the equations defining these operators. Let $\mathbf{x}$, $\mathbf{y}$, $\mathbf{z}$ be an orthonormal basis of $\mathbf{R}^3$ that, by stipulation, is right-handed. Let $\sigma_x$, $\sigma_y$, $\sigma_z$ be operators associating the vectors $\mathbf{x}$, $\mathbf{y}$, $\mathbf{z}$ with values $\pm 1$. Then, QM defines these operators by the equation:

$$\sigma_i \sigma_j + \sigma_j \sigma_i = 2 \delta_{ij} \mathbf{1}, \qquad (1)$$

where i, j = x, y, z and $\mathbf{1}$ is the unit operator. If spin operators for more than one system appear, they are distinguished by superscripts 1, 2, 3, … and QM prescribes, for a, b, … = 1, 2, 3, … and a ≠ b, for any i, j that:

$$\sigma^a_i \sigma^b_j = \sigma^b_j \sigma^a_i. \qquad (2)$$

We now recall two well-known no-hidden-variables arguments. They operate with systems consisting of two or three spin-½ particles. We refer to the systems measured for certain observables simply as *systems*. E.g., the Peres-Mermin (PM) system is a two-particle spin-½ system measured for the following nine observables: $\sigma^1_x$, $\sigma^2_x$, $\sigma^1_y$, $\sigma^2_y$, $\sigma^1_x \sigma^2_x$, $\sigma^1_y \sigma^2_y$, $\sigma^1_x \sigma^2_y$, $\sigma^1_y \sigma^2_x$, $\sigma^1_z \sigma^2_z$. (See [4, 5].) From (1) and (2), it follows that these observables satisfy the following six constraints (where $\mathbf{1}$ is the unit observable):

$$\sigma^1_y \sigma^2_y \sigma^1_y \sigma^2_y = \mathbf{1}, \qquad (3b)$$
$$\sigma^1_x \sigma^2_y \sigma^1_x \sigma^2_y = \mathbf{1}, \qquad (3c)$$
$$\sigma^1_y \sigma^2_x \sigma^1_y \sigma^2_x = \mathbf{1}, \qquad (3d)$$
$$\sigma^1_x \sigma^2_y \sigma^1_y \sigma^2_x \sigma^1_z \sigma^2_z = \mathbf{1}, \qquad (3e)$$
$$\sigma^1_x \sigma^2_x \sigma^1_y \sigma^2_y \sigma^1_z \sigma^2_z = -\mathbf{1}. \qquad (3f)$$

Assume now what is called the Product Rule, i.e. the claim that for any two commuting observables **A** and **B**, the value of their product, written as [**AB**], equals the product of their values, written as [**A**] [**B**]. (This rule follows when we require that the algebraic relations among the operators or observables are mirrored in the values [25]). Given this assumption, equations (3a-f) lead to constraints for the values of the nine observables, namely:

$$[\sigma^1_x \sigma^2_x] [\sigma^1_x] [\sigma^2_x] = 1, \qquad (4a)$$
$$[\sigma^1_y \sigma^2_y] [\sigma^1_y] [\sigma^2_y] = 1, \qquad (4b)$$
$$[\sigma^1_x \sigma^2_y] [\sigma^1_x] [\sigma^2_y] = 1, \qquad (4c)$$
$$[\sigma^1_y \sigma^2_x] [\sigma^1_y] [\sigma^2_x] = 1, \qquad (4d)$$
$$[\sigma^1_x \sigma^2_y] [\sigma^1_y \sigma^2_x] [\sigma^1_z \sigma^2_z] = 1, \qquad (4e)$$
$$[\sigma^1_x \sigma^2_x] [\sigma^1_y \sigma^2_y] [\sigma^1_z \sigma^2_z] = -1. \qquad (4f)$$

So, QM ((1) and (2) above) and the Product Rule jointly predict that the result of measuring the three observables in any line (4a-f) meets the constraint explicated in that line. On the other hand, given an assumption of faithful measurement, any such measurement of three observables just reveals their pre-existing values. As a consequence, any of the values appearing on the left of (4) must be the same in both places; this is the non-contextuality assumption. So, all six constraints in (4) must jointly be met. But this is impossible because the product of all the left sides of (4) equals 1, while the one of all the right sides equals –1. Since this argument operates with the PM system, we call it the *PM argument*.

The *Greenberger-Horne-Zeilinger (GHZ) argument* ([6]) can be cast in a similar form ([5]). The GHZ system is a three-particle spin-½ system measured for the following ten observables: $\sigma^1_x$, $\sigma^1_y$, $\sigma^2_x$, $\sigma^2_y$, $\sigma^3_x$, $\sigma^3_y$, $\sigma^1_x \sigma^2_y \sigma^3_y$, $\sigma^1_y \sigma^2_x \sigma^3_y$, $\sigma^1_y \sigma^2_y \sigma^3_x$, $\sigma^1_x \sigma^2_x \sigma^3_x$. For these observables, (1) and (2) yield these five constraints:

$$\sigma^1_x \sigma^2_y \sigma^3_y \sigma^1_x \sigma^2_y \sigma^3_y = 1, \quad (5a)$$
$$\sigma^1_y \sigma^2_x \sigma^3_y \sigma^1_y \sigma^2_x \sigma^3_y = 1, \quad (5b)$$
$$\sigma^1_y \sigma^2_y \sigma^3_x \sigma^1_y \sigma^2_y \sigma^1_x = 1, \quad (5c)$$
$$\sigma^1_x \sigma^2_x \sigma^3_x \sigma^1_x \sigma^2_x \sigma^1_x = 1, \quad (5d)$$
$$\sigma^1_x \sigma^2_x \sigma^3_x \sigma^1_x \sigma^2_y \sigma^3_y \sigma^1_y \sigma^2_x \sigma^3_y \sigma^1_y \sigma^2_y \sigma^3_x = -1. \quad (5e)$$

Assuming the Product Rule again, the values of these observables must satisfy these five constraints:

$$[\sigma^1_x \sigma^2_y \sigma^3_y] [\sigma^1_x] [\sigma^2_y] [\sigma^3_y] = 1, \quad (6a)$$
$$[\sigma^1_y \sigma^2_x \sigma^3_y] [\sigma^1_y] [\sigma^2_x] [\sigma^3_y] = 1, \quad (6b)$$
$$[\sigma^1_y \sigma^2_y \sigma^3_x] [\sigma^1_y] [\sigma^2_y] [\sigma^1_x] = 1, \quad (6c)$$
$$[\sigma^1_x \sigma^2_x \sigma^3_x] [\sigma^1_x] [\sigma^2_x] [\sigma^1_x] = 1, \quad (6d)$$
$$[\sigma^1_x \sigma^2_x \sigma^3_x] [\sigma^1_x \sigma^2_y \sigma^3_y] [\sigma^1_y \sigma^2_x \sigma^3_y] [\sigma^1_y \sigma^2_y \sigma^3_x] = -1. \quad (6e)$$

Again, each line contains a set of three mutually commuting observables, i.e. QM plus the Product Rule predicts that the result of measuring the three observables in any line meets the constraint explicated in that line. Again, we make an assumption of faithful measurement; we approach the GHZ system picking an arbitrary set of three observables (an arbitrary one of lines (5a-e)) and measurement then faithfully reveals pre-existing values of the observables. In particular, any of the values appearing on the left of (5) must be the same in both places; this is the non-contextuality assumption, again. So, the five constraints in (5a-e) must jointly be met. Again, this is impossible since the product of all the left sides is 1, but the one of all the right sides is – 1.

This argument using the GHZ system and a non-contextuality assumption can also be framed as a non-locality argument that may be referred to as the *Bell-GHZ argument*. For future reference, we directly quote the argument from Mermin:

> [Consider] a system in a particular one of the simultaneous eigenstates of the three operators $\sigma^1_x \sigma^2_y \sigma^3_y$, $\sigma^1_y \sigma^2_x \sigma^3_y$, and $\sigma^1_y \sigma^2_y \sigma^3_x$ – say the state $\Phi$ in which the three have eigenvalue 1. It follows that $\Phi$ is also an eigenstate of $\sigma^1_x \sigma^2_x \sigma^3_x = - (\sigma^1_x \sigma^2_y \sigma^3_y) (\sigma^1_y \sigma^2_x \sigma^3_y) (\sigma^1_y \sigma^2_y \sigma^3_x)$ with eigenvalue – 1. We now note that if three mutually well separated particles have their spins in this state, then we can learn in advance the result […of measuring any component from the result of measuring the other two …], since the product of all three measurements in the state $\Phi$ must be unity. [… Assuming locality, one is thus] impelled to conclude that the results of measuring either component of any of the three particles must have already been specified prior to any of the measurements – i.e. that any particular system in the state $\Phi$ must be characterized by numbers $m_x^1$, $m_y^1$, $m_x^2$, $m_y^2$, $m_x^3$, $m_y^3$ which specify the results of whichever of the four different sets (xyy, yxy, yyx, xxx) of three single particle spin measurements one might choose to make on the three far apart particles. Because, however, $\Phi$ is an eigenstate of $\sigma^1_x \sigma^2_y \sigma^3_y$, $\sigma^1_y \sigma^2_x \sigma^3_y$, $\sigma^1_y \sigma^2_y \sigma^3_x$, and $\sigma^1_x \sigma^2_x \sigma^3_x$ with respective eigenvalues 1, 1, 1, and – 1, the products of the four trios of 1's or – 1's must satisfy the relations

$$m_x^1 m_y^2 m_y^3 = 1, \quad (7a)$$
$$m_y^1 m_x^2 m_y^3 = 1, \quad (7b)$$
$$m_y^1 m_y^2 m_x^3 = 1, \quad (7c)$$
$$m_x^1 m_x^2 m_x^3 = -1, \quad (7d)$$

> which, once again, are mutually inconsistent, the product of the four left sides being necessarily +1.' ([26])

## III. INTERPRETATION OF THE GEOMETRIC ALGEBRA OF R³

In the following, our goal is to show how (4a-f), (6a-g), and (7a-d), instead of each implying the falsity 1 = – 1, can each lead to a *truth*. This will indeed be done for (4a-f), (6a-g), and a proxy for (7a-d), but only given a substantial reinterpretation of the formalism. The basic idea is to reinterpret the values of observables $\sigma_x$, $\sigma_y$, $\sigma_z$, conventionally conceived to be scalars, as vectors. This will avoid the false equation 1 = – 1.

To prepare for this reinterpretation, we recall basic ideas of geometric algebra (GA), the mathematical theory exploring spaces of multivectors generated from vector spaces by means of the geometric product. In particular, we consider the multivector space **G³** that is generated from the vector space **R³**, also called the geometric algebra **G³**, which is described in detail in many places in the literature ([27, 28]). Let {$e_1$, $e_2$, $e_3$} be an orthonormal and right-handed basis of **R³**. Given the geometric product, the vectors $e_1$, $e_2$, $e_3$ instantiate the very structure that was used in (1) to define the Pauli operators, i.e.:

$$e_i e_j + e_j e_i = 2 \delta_{ij}, \quad (8)$$

for i, j = 1, 2, 3, which yields:

$$e_i e_i = 1 \quad \text{and} \quad (9)$$

$$e_i e_j = - e_j e_i, \quad \text{for } i \neq j. \quad (10)$$

Note that (8) is not a stipulation but follows from the properties of our basis {$e_1$, $e_2$, $e_3$} and the definition of the geometric product on **R³** [29]. Important multivectors (elements of **G³**) constituted by the basis vectors with the geometric product are bi- and trivectors. The product $e_i e_j$, for $i \neq j$, is called a unit bivector; the product $e_i e_j e_k$, for $i \neq j$, $j \neq k$, $i \neq k$, is called a unit trivector.

### A. Interpreting bivectors

Bi- and trivectors can be interpreted geometrically. We consider bivectors first. A bivector $e_i e_j$ can be interpreted (A) as a rotation operator or (B) as an orientation in the i,j-plane. We consider the implications of both interpretations in parallel. (A) Since $e_i e_j e_i = -e_j$ and $e_i e_j e_j = e_i$, the product $e_i e_j$ may be interpreted as a rotation operator acting on vectors in the i,j-plane and rotating them by $\pi/2$. (This interpretation is well-known in the GA literature ([30]).) Assuming that $e_i e_j = e_i e_j \varepsilon_{ijk}$ and $e_i e_j$ acts on a vector by being left-multiplied to it, the effected rotation is *clockwise*. (B) Alternatively, $e_i e_j$ may be interpreted not as an operator acting on vectors, but as an orientation in the i,j-plane of some system extended in that plane (see Appendix A1 for further discussion). Assuming again that (i, j, k) is an even permutation of (1, 2, 3), the bivector $e_i e_j$ is a *counter-clockwise* orientation, because, viewing the i,j-plane from the tip of $e_k$, $e_j$ is at an angle of $\pi/2$ to the left of $e_i$. The clockwise orientations in the i,j-plane are of the form $e_i e_j \varepsilon_{jik}$, the counter-clockwise ones of the form $e_i e_j \varepsilon_{ijk}$.

Both interpretations, (A) and (B), afford an understanding of (9) and (10). Reference to the i,j-plane presupposes reference to vectors $e_i$, $e_j$ with $i \neq j$. But there are no such vectors in (9), hence, in (9) there is no reference to any entity presupposing the i,j-plane. By contrast, in (10) there is a reference to (A) a rotation operator or (B) an orientation $e_i e_j$ in the i,j-plane. (10) then claims that this entity $e_i e_j$ (in (A): an operator, in (B): an orientation) can be constituted by different components: the vectors $e_i$, $e_j$, multiplied in this order, constitute $e_i e_j$, and so do the vectors $(-e_j)$, $e_i$, multiplied in this order. We can speak of bivectors being constituting by vectors. Thus, the four bivectors $e_i e_j$ and $(-e_j) e_i$ and $(-e_i)(-e_j)$ and $e_j(-e_i)$ are constituted by *different* vectors (multiplied in a certain order) and yet are the *same* bivector, interpretable as either (A) a rotation operator for a clockwise $\pi/2$-rotation in the i,j-plane or (B) a counter-clockwise orientation in that plane (see Appendix A1 for further discussion).

From (9), we can derive:

$$e_i e_j e_j e_i = 1. \qquad (11)$$

Similarly, from (9) and (10), we can derive:

$$e_i e_j e_i e_j = -1. \qquad (12)$$

Again, these equations allow different geometric interpretations. (A) Assuming that $e_i e_j$ is an operator inducing a clockwise $\pi/2$-rotation, $e_j e_i$ is its counterpart: an operator for a counter-clockwise $\pi/2$-rotation. Accordingly, the sequence of both, $e_i e_j e_j e_i$, is an operator for a counter-clockwise $\pi/2$-rotation followed by a clockwise $\pi/2$-rotation, which is the identity operator, thus affording an interpretation of (11); similarly, the sequence $e_i e_j e_i e_j$ is an operator for two consecutive clockwise $\pi/2$-rotations, which is an operator effecting a clockwise $\pi$-rotation, thereby changing the sign of every vector in the i,j-plane and thus affording an interpretation of (12). This interpretation of unit bivectors as rotation operators, however, is dissatisfying because it presupposes the existence of a vector that is rotated by means of a rotation operator; accordingly, without specifying such a vector the operators on the left of (11) and (12) are devoid of geometric content.

(B) An alternative interpretation of (11) and (12) can be given when we ignore rotations and focus on $e_i e_j$ as an orientation of some system in the i,j-plane. The vectors $e_i$ and $e_j$, when multiplied in this order, can be viewed as thereby *generating* the bivector $e_i e_j$, which we have identified with a counter-clockwise orientation. Accordingly, a clockwise orientation, if right-multiplied with $e_i e_j$, can be viewed as *annihilating* it. To define this idea, we stipulate that an orientation annihilates an orientation iff their product equals 1. Since there are exactly two different orientations $e_i e_j$ (counter-clockwise) and $e_j e_i$ (clockwise) in the i,j-plane and they multiply to 1, these two orientations annihilate each other, which is the content of (9). Since $e_i e_j$ and $(-e_j) e_i$ are the same orientation they do not annihilate each other and their product equals $-1$, which, given (10), is the content of (12).

### B. Interpreting trivectors

We turn to the interpretation of trivectors. We begin with the trivial identity:

$$e_i e_j e_k = e_i e_j e_k, \qquad (13)$$

where (i, j, k) is any permutation of (1, 2, 3). From (13), by (9) and (10):

$$e_i e_j e_k e_k e_j e_i = 1, \qquad (14)$$

$$e_i e_j e_k e_i e_j e_k = -1. \qquad (15)$$

(14) and (15) again have competing interpretations, now depending on whether $e_i e_j e_k$ and its permutations are interpreted (A) as operators acting on vectors or bivectors or (B) as 3D orientations.

(A) First, $e_i e_j e_k$ can be interpreted as an operator mapping vectors onto their dual bivectors and bivectors onto their dual vectors. Then, (14) and (15) can be viewed as illustrating that these pairs of mappings do not exhibit a self-duality, as, for any vector $\mathbf{v} \in \mathbf{R}^3$, $(e_i e_j e_k e_i e_j e_k) \mathbf{v} = -\mathbf{v}$. Geometrically, $(e_i e_j e_k e_i e_j e_k)$ is an inversion through the origin of $\mathbf{R}^3$, while $e_i e_j e_k e_k e_j e_i$ is just the identity. Again, this interpretation is dissatisfying, as it requires the choice of a vector (alternatively, now, a bivector) and does not suggest an obvious geometric meaning on its own.

(B) Alternatively, $e_i\, e_j\, e_k$ can be interpreted as an orientation of a system extended not only in the i,j-plane but in all three dimensions of $\mathbf{R}^3$. We begin by assuming that (i, j, k) is an even permutation of (1, 2, 3) such that (since we have assumed $e_1\, e_2\, e_3$ to be right-handed) $e_i\, e_j\, e_k$ is right-handed and $e_k\, e_j\, e_i$ is left-handed. It is easy to see that 3D orientations are related just as 2D ones. The vectors $e_i$, $e_j$ and $e_k$, when multiplied in this order, thereby generate the trivector $e_i\, e_j\, e_k$, which we have identified with a right-handed orientation. A second orientation, if right-multiplied with the orientation $e_i\, e_j\, e_k$, can be viewed as annihilating it, as before. Again, we stipulate that an orientation annihilates an orientation iff their product equals 1. Since there are exactly two different orientations $e_i\, e_j\, e_k$ (right-handed) and $e_j\, e_i\, e_k$ (left-handed) and their product equals 1, they annihilate each other, which, given (10), is the content of (14). Since $e_i\, e_j\, e_k$ and $(-e_j)\, e_i\, e_k$ are the same orientation they do not annihilate each other and their product equals $-1$, which, given (8), is the content of (15). More directly, $e_i\, e_j\, e_k$ is identical with itself and does not annihilate itself. Thus, its product with itself does not equal 1 but $-1$, which again is the content of (15).

### C. Vector variables

So far, we have considered our two interpretations of bi- and trivectors (A) and (B) pari passu. In the following, we will drop the operator interpretation (A) and retain only the orientation interpretation (B) in order to further explore the representation of orientations in GA. We will now generalize interpretation (B) in two directions: first, from vectors to vector variables and second, from one to several bases of $\mathbf{R}^3$.

We begin with the variables. The structure of (11) and (12) suggests the introduction of vector-valued variables. We introduce variables $\sigma_i$, $\sigma_j$, $\sigma_k$, that are two-valued, i.e. can take on values $\pm e_i$, $\pm e_j$, $\pm e_k$, respectively. In (9), $e_i$ can be replaced with $-e_i$, ad libitum, as long as we do it in both occurrences, and similarly for $e_j$; hence, (11) and (12) can be generalized to:

$$\sigma_i\, \sigma_j\, \sigma_j\, \sigma_i = 1, \qquad (16)$$

$$\sigma_i\, \sigma_j\, \sigma_i\, \sigma_j = -1. \qquad (17)$$

(16) and (17) have the same interpretation as (11) and (12), i.e. one orientation in the i,j-plane annihilates the other but does not annihilate itself.

A similar generalization suggests itself for the 3D case. In (14) and (15) $e_i$, $e_j$, and $e_k$ can be freely exchanged with their negatives without affecting the right sides, thus we can again generalize them to the $\sigma$-variables:

$$\sigma_i\, \sigma_j\, \sigma_k\, \sigma_k\, \sigma_j\, \sigma_i = 1, \qquad (18)$$

$$\sigma_i\, \sigma_j\, \sigma_k\, \sigma_i\, \sigma_j\, \sigma_k = -1. \qquad (19)$$

(18) expresses that $\sigma_i\, \sigma_j\, \sigma_k = \sigma_i\, \sigma_j\, \sigma_k$ and (19) expresses that $\sigma_i\, \sigma_j\, \sigma_k \neq \sigma_k\, \sigma_j\, \sigma_i$. Since there are only two orientations corresponding to odd and even permutations of (1, 2, 3), the latter entails $\sigma_i\, \sigma_j\, \sigma_k = -\sigma_k\, \sigma_j\, \sigma_i$. (18) and (19) have the same interpretation as (14) and (15): different orientations in $\mathbf{R}^3$ annihilate each other and no orientation annihilates itself. Note that the vector value equations, and a fortiori the vector variable equations, are compatible with any choice of orientation in the i,j-plane or $\mathbf{R}^3$. The constraints that arise refer to different and identical orientations but the actual orientation of a system satisfying these equations does not have to be specified.

### D. Identities of orientations

So far, we have considered orientations constituted by elements of the basis $\{e_1, e_2, e_3\}$ – what might be called **e**-orientations. Equation (10) expressed that certain two **e**-orientations in the i,j-plane are identical and similarly (13) expresses that certain two **e**-orientations in $\mathbf{R}^3$ are identical. Consider now orientations constituted by vectors from different bases. More explicitly, consider $\{e_1, e_2, e_3\}$ and a second orthonormal basis $\{f_1, f_2, f_3\}$ (from now on, the elements of these bases are often briefly called the **e**'s and the **f**'s). Initially, we leave open whether or not the products $e_i\, e_j\, e_k$ and $f_l\, f_m\, f_n$ (where (l, m, n) is any permutation of (1, 2, 3)) are the same orientation. Now assume that these two orientations are in fact identical, i.e. assume:

$$e_i\, e_j\, e_k = f_l\, f_m\, f_n. \qquad (20)$$

Without loss of generality (see Appendix A4), we can rewrite (20) as:

$$e_i\, e_j\, e_k = f_i\, f_j\, f_k. \qquad (21)$$

Since the **f**'s are orthonormal, an analogue of (8) holds for them. Hence, (21) is equivalent to:

$$e_i\, e_j\, e_k\, f_j\, f_i\, f_k = 1 \quad \text{and} \qquad (22)$$

$$e_i\, e_j\, e_k\, f_i\, f_j\, f_k = -1. \qquad (23)$$

Now, what does it mean to say that two orientations are identical? We have interpreted unit bi- and trivectors not as orientations characterizing a certain plane in $\mathbf{R}^3$ or the whole of $\mathbf{R}^3$, but as orientations characterizing *systems* extended in a plane or in $\mathbf{R}^3$. Hence, we can distinguish two kinds of the identity of orientations: one where two orientations of the same system are identical and another where two orientations of different systems are identical. The distinction is intuitively accessible, though conceptually non-trivial (see Appendix A4 for further discussion).

Assume that a system in the i,j-plane or in **R³** has a unique (2D or 3D) orientation. Then the two kinds of identity just considered lead to different consequences. It suffices to consider the 3D case. Consider first two orientations, one constituted by **e**'s, the other by **f**'s, and pertaining to the same system. Since the system's orientation is unique, the two orientations must be identical, i.e. (21) is true by construction. Since the **e**'s are a basis of **R³**, the **f**'s can be written in terms of the **e**'s and it is obvious that not all **e**'s and **f**'s commute.

Consider second two orientations, one constituted by **e**'s, the other by **f**'s, and pertaining to two different systems. In this case, (21) is contingently true or false. Given the **f**'s, their order is arbitrary such that the orientation they constitute is arbitrary. Hence, given an orientation $\mathbf{e}_i \mathbf{e}_j \mathbf{e}_k$ of one system, and a basis of **f**'s chosen to describe a second system's orientation, the order of **f**'s and hence this second orientation is arbitrary. For this case, we would like to assume that the orientations' constituents, the **e**'s and **f**'s, all *commute* – but will be able to do so only with a certain qualification.

The distinction of **e**'s and **f**'s with respect to one system turns out to be superfluous. The orientation $\mathbf{f}_i \mathbf{f}_j \mathbf{f}_k$ can be written in terms of the **e**'s and is identical with $\mathbf{e}_i \mathbf{e}_j \mathbf{e}_k$ iff (21) is true. Given our assumptions that $\mathbf{e}_i \mathbf{e}_j \mathbf{e}_k$ and $\mathbf{f}_i \mathbf{f}_j \mathbf{f}_k$ are orientations of the same system and this system has a unique orientation, (21) is necessarily true, i.e. $\mathbf{e}_i \mathbf{e}_j \mathbf{e}_k = \mathbf{f}_i \mathbf{f}_j \mathbf{f}_k$, as a matter of logic. Hence, from now on we stop to refer to components of orientations of the *same* system via different sets of vectors, the **e**'s and **f**'s, and reserve the letters '**e**' and '**f**' for constituents of orientations of *different* systems.

We want to assume that these constituents, the **e**'s and **f**'s, generally commute but have to allow for one qualification. The **e**'s and **f**'s cannot be assumed to commute in the presence of identities between individual **e**'s and **f**'s. Assume (21) and assume, in addition, that $\mathbf{e}_k = \mathbf{f}_k$, whence it follows that $\mathbf{e}_i \mathbf{e}_j = \mathbf{f}_i \mathbf{f}_j$. It is easy to show that in this case the **e**'s and **f**'s cannot all commute (see Appendix A3 for a proof). So, what we finally assume is that all the **e**'s and **f**'s commute iff no identities between any of the **e**'s and **f**'s obtain. Below, we will consider also a third orthonormal basis {$\mathbf{g}_1, \mathbf{g}_2, \mathbf{g}_3$}, the **g**'s. Qualifications analogous to the ones for **e**'s and **f**'s hold also for the **e**'s and **g**'s and for the **f**'s and **g**'s.

In (22) and (23), we cannot replace an arbitrary component vector with its negative without falsifying the equations, but we can replace any *two* components with their negatives, ad libitum. (22) and (23) can thus again be generalized by means of σ-variables, but only given a condition. We introduce variables $\sigma^1_i, \sigma^1_j, \sigma^1_k$, with possible values $\pm \mathbf{e}_i, \pm \mathbf{e}_j, \pm \mathbf{e}_k$, respectively, and variables $\sigma^2_i, \sigma^2_j, \sigma^2_k$, with possible values $\pm \mathbf{f}_i, \pm \mathbf{f}_j, \pm \mathbf{f}_k$, respectively. The variables' values must be chosen so as to satisfy (21), i.e. so that an even number of variables changes its value. Given this condition, (22) and (23) can be generalized to:

$$\sigma^1_i \sigma^1_j \sigma^1_k \sigma^2_j \sigma^2_i \sigma^2_k = 1 \text{ and} \quad (24)$$

$$\sigma^1_i \sigma^1_j \sigma^1_k \sigma^2_i \sigma^2_j \sigma^2_k = -1. \quad (25)$$

(24) and (25) complement (18) and (19) for the case of orientations of different systems.

We assume that there are no identities between the **e**'s and the **f**'s and thus may assume that they commute with each other. Since they commute, so do the ($\sigma^1$)'s with the ($\sigma^2$)'s and we can rewrite (24) and (25) as:

$$\sigma^1_i \sigma^2_j \sigma^1_j \sigma^2_i \sigma^1_k \sigma^2_k = 1, \quad (26)$$

$$\sigma^1_i \sigma^2_i \sigma^1_j \sigma^2_j \sigma^1_k \sigma^2_k = -1. \quad (27)$$

We have considered the identity of 3D orientations pertaining to different systems and now turn to the 2D case. Consider the **e**'s and **f**'s again, plus a third orthonormal basis {$\mathbf{g}_1, \mathbf{g}_2, \mathbf{g}_3$}, the **g**'s. Since they all obey (8), they also all obey (11) and (12), i.e. $\mathbf{e}_i \mathbf{e}_i \mathbf{e}_j \mathbf{e}_j = -(\mathbf{f}_i \mathbf{f}_j \mathbf{f}_i \mathbf{f}_j) = \mathbf{g}_i \mathbf{g}_i \mathbf{g}_j \mathbf{g}_i = 1$, which yield:

$$\mathbf{e}_i \mathbf{e}_i \mathbf{e}_j \mathbf{e}_j \mathbf{f}_i \mathbf{f}_j \mathbf{f}_i \mathbf{f}_j \mathbf{g}_i \mathbf{g}_i \mathbf{g}_j \mathbf{g}_i = -1. \quad (28)$$

We assume that there are no identities between the **e**'s, **f**'s, and **g**'s and thus may assume that they all commute. Hence, from (28):

$$\mathbf{e}_i \mathbf{f}_i \mathbf{g}_i \mathbf{e}_i \mathbf{f}_j \mathbf{g}_j \mathbf{e}_j \mathbf{f}_i \mathbf{g}_i \mathbf{e}_j \mathbf{f}_j \mathbf{g}_i = -1. \quad (29)$$

Since any one of the **e**'s, **f**'s, and **g**'s appearing in (28) appears twice, we can generalize (28) to σ-variables as:

$$\sigma^1_i \sigma^1_i \sigma^1_j \sigma^1_j \sigma^2_i \sigma^2_j \sigma^2_i \sigma^2_j \sigma^3_i \sigma^3_j \sigma^3_j \sigma^3_i = -1. \quad (30)$$

Since the **e**'s, **f**'s and **g**'s all mutually commute, so do the ($\sigma^1$)'s, ($\sigma^2$)'s and ($\sigma^3$)'s. Thus, from (29) by generalization to σ-variables or, alternatively, from (30) by commutation of the σ's pertaining to different systems:

$$\sigma^1_i \sigma^2_i \sigma^3_i \sigma^1_i \sigma^2_j \sigma^3_j \sigma^1_j \sigma^2_i \sigma^3_j \sigma^1_j \sigma^2_j \sigma^3_i = -1. \quad (31)$$

In our considerations of orientations pertaining to two or three different systems, we have explicitly assumed that there are no identities between the **e**'s, **f**'s, and **g**'s. We now finally drop this assumption and consider a triple of orientations with the additional property of identical components (see the discussion in Appendix A4). As emphasized above, this excludes the claim that the components of different systems' orientations commute. We consider the **e**'s, **f**'s and **g**'s lying in the i,j-plane and their products; in particular, we are interested in these four multivectors: $\mathbf{e}_i \mathbf{f}_j \mathbf{g}_j, \mathbf{e}_j \mathbf{f}_i \mathbf{g}_j, \mathbf{e}_j \mathbf{f}_j \mathbf{g}_i, \mathbf{e}_i \mathbf{f}_i \mathbf{g}_i$. We choose the most natural identities: $\mathbf{e}_i = \mathbf{f}_i = \mathbf{g}_i$ and $\mathbf{e}_j = \mathbf{f}_j = \mathbf{g}_j$. From this assumption and using (9) and (10), we immediately get:

|   |   |   |   |   |   |
|---|---|---|---|---|---|
| $e_i\, f_j\, g_j$ | = | $e_i\, e_j\, e_j$ | = | $e_i$ | (32a) |
| $e_j\, f_i\, g_j$ | = | $e_j\, e_i\, e_j$ | = | $-e_i$ | (32b) |
| $e_j\, f_j\, g_i$ | = | $e_j\, e_j\, e_i$ | = | $e_i$ | (32c) |
| $e_i\, f_i\, g_i$ | = | $e_i\, e_i\, e_i$ | = | $e_i.$ | (32d) |

We note that all three columns in (32a-d) multiply to $-1$, in contrast with the similar, yet inconsistent, system of equations (7a-d).

## IV. DERIVATION OF THE PM AND GHZ VALUE CONSTRAINTS

Evidently, our equations (26), (27), (30) and (32) are structurally similar to the equations (3e), (3f), (5e) and (7) characterizing the PM and GHZ systems. In deriving the former, we have made use of the idea that bi- and trivectors are orientations attached to systems that are extended in two or three spatial dimensions. This suggests an interpretation of the examples in terms of such orientations.

Our hidden variable model consists solely in the introduction of an orientation characterizing individually each subsystem of the PM and GHZ systems. More specifically, we replace the values of the spin operators by components of orientations, either vectors or their geometric products, and rule that these components obey the GA equations ((8) and its implications) for such multivectors. Formally, we replace the quantum-mechanical *operators*, which are scalar-valued, by vector *variables*, which are vector-valued, but we indiscriminately write both (the operators and the variables) as $\sigma$'s; moreover, we write the values of the variables in square brackets, as we did with the ones of the operators. For the elementary vector variables, consisting of only one of the $\sigma$'s, we assume that $[\sigma^1_x] = \pm e_1$, $[\sigma^1_y] = \pm e_2$, $[\sigma^1_z] = \pm e_3$; similarly $[\sigma^2_x] = \pm f_1$, $[\sigma^2_y] = \pm f_2$, $[\sigma^2_z] = \pm f_3$; and $[\sigma^3_x] = \pm g_1$, $[\sigma^3_y] = \pm g_2$, $[\sigma^3_z] = \pm g_3$. For simplicity, we set all these values to positive vectors ($[\sigma^1_x] = e_1$, and so on), since, as will become apparent, the choice of these values, and thus the choice of one particular orientation, is arbitrary. Call this ascription of exclusively positive vectors to the elementary $\sigma$-variables the *value assumption*. Concerning the products of the elementary variables, we assume a counterpart of the Product Rule, i.e. if **A**, **B** are commuting variables, the value $[AB]$ of their product is given by $[AB] = [A][B]$. Call this the *product assumption*. (It is the Product Rule from sect. 2, adapted to vector variables with the interpretation that the function [...] can be applied to vectors and their products.) Finally, assume the commutativity of **e**'s, **f**'s, and **g**'s and call it the commutativity assumption or briefly: *commutativity*. Given these three assumptions, the PM and GHZ value constraints, i.e. equation systems (4) and (6) above, are readily derived from GA. We begin with the PM system. (4a) above is derived as follows:

|   |   |   |
|---|---|---|
| $[\sigma^1_x\, \sigma^2_x]\, [\sigma^1_x]\, [\sigma^2_x]$ | | (33) |
| = | $e_1\, f_1\, e_1\, f_1$ | |
| = | $e_1\, e_1\, f_1\, f_1$ | |
| = | 1 | |

In (33), equation 1 is due to the value and product assumptions, equation 2 to the commutativity of **e**'s and **f**'s, and equation 3 to GA, more specifically (9) above. Equations (4b-d) follow on the same lines. We can derive (4e) as follows:

|   |   |   |
|---|---|---|
| $[\sigma^1_x\, \sigma^2_y]\, [\sigma^1_y\, \sigma^2_x]\, [\sigma^1_z\, \sigma^2_z]$ | | (34) |
| = | $e_1\, f_2\quad e_2\, f_1\quad e_3\, f_3$ | |
| = | $e_1\, e_2\, e_3\quad f_2\, f_1\, f_3$ | |
| = | 1. | |

The equations in (34) again hold due to the value and product assumptions, commutativity of **e**'s and **f**'s, and finally GA, more specifically (22) above. (4f) is obtained in the same way, using (23). The heart of each one of these simple derivations is its last step. The last equation of (34), i.e.: $e_1\, e_2\, e_3\, f_2\, f_1\, f_3 = 1$, is an instance of (22) and in a parallel derivation of (4f) the last equation would be an instance of (23): $e_1\, e_2\, e_3\, f_1\, f_2\, f_3 = -1$. Multiplication of all value equations in the PM argument boils down to multiplying these two instances, giving us (with the help of commutativity):

$$e_1\, e_2\, e_3\, e_1\, e_2\, e_3\, f_2\, f_1\, f_3\, f_1\, f_2\, f_3 = -1. \qquad (35)$$

Clearly, (35) could not be satisfied if the **e**'s and **f**'s were real numbers, but since we assume them to be vectors, obeying (8) and its implications, (35) follows directly (from (22), (23) and commutativity).

We can treat the GHZ system in the same way. We derive (6a) as follows:

|   |   |   |
|---|---|---|
| $[\sigma^1_x\, \sigma^2_y\, \sigma^3_y]\, [\sigma^1_x]\, [\sigma^2_y]\, [\sigma^3_y]$ | | (36) |
| = | $e_1\, f_2\, g_2\, e_1\, f_2\, g_2$ | |
| = | $e_1\, e_1\quad f_2\, f_2\quad g_2\, g_2$ | |
| = | 1. | |

Here, we have used the value assumption, the commutativity of **e**'s, **f**'s, and **g**'s and GA, namely (8) above. Similarly, for (6b-d). Finally, for (6e):

|   |   |   |
|---|---|---|
| $[\sigma^1_x\, \sigma^2_x\, \sigma^3_x]\, [\sigma^1_x\, \sigma^2_y\, \sigma^3_y]\, [\sigma^1_y\, \sigma^2_x\, \sigma^3_y]\, [\sigma^1_y\, \sigma^2_y\, \sigma^3_x]$ | | (37) |
| = | $e_1\, f_1\, g_1\quad e_1\, f_2\, g_2\quad e_2\, f_1\, g_2\quad e_2\, f_2\, g_1$ | |
| = | $e_1\, e_1\, e_2\, e_2\quad f_1\, f_2\, f_1\, f_2\quad g_1\, g_2\, g_2\, g_1$ | |
| = | $-1$ | |

Again, we have used the value assumption, commutativity, and GA ((8) above).

Our derivations have not required an assumption of non-contextuality, but obviously such an assumption is respected. We have effectively assumed that in (4a-f) all value expressions in square brackets are vectors or their products, but not scalars, and we have arbitrarily restricted ourselves to vectors with a positive sign. These vectors are the same for every occurrence of any vector variable. E.g., we have $[\sigma^1_x] = \mathbf{e}_1$ in both (4a) and (4c) and $[\sigma^1_x \sigma^2_y] = \mathbf{e}_1 \mathbf{f}_2$ in both (4c) and (4e), and similarly for all other expressions on the left of (4). The values of all variables of the PM system are the same in different contexts, i.e. across different lines of (4), and thus are manifestly non-contextual. The same holds for (6a-e) and the variables of the GHZ system.

## V. DERIVATION OF A PROXY FOR THE BELL-GHZ CONSTRAINTS

Defusing the Bell-GHZ argument is less straightforward. Mermin's version of the argument ends with the false equation $1 = -1$, as did the PM and GHZ arguments, but Mermin's m's are assumed to individually equal $\pm 1$, i.e. they are real numbers, which forbids to conceive of them as vectors. However, we can drop Mermin's assumption in line with our basic idea to re-interpret the values of QM observables as vectors; while vectors cannot be replaced by real numbers without destroying the former's anti-commutation properties, some vector products are real numbers. Our proxy for (7) thus is the system of four equations given by the third and fourth columns of (38) below. Multiplying the four equations and using (9) and (10), we get $-1 = -1$, a trivial truth replacing the falsity $1 = -1$.

We derive (38) as follows. Consider again the vector variables $\sigma^1_x, \sigma^1_y, \sigma^2_x, \sigma^2_y, \sigma^3_x, \sigma^3_y$ and, again for simplicity, let them all have positive vector values: $[\sigma^1_x] = \mathbf{e}_1$, $[\sigma^1_y] = \mathbf{e}_2$, $[\sigma^2_x] = \mathbf{f}_1$, $[\sigma^2_y] = \mathbf{f}_2$, $[\sigma^3_x] = \mathbf{g}_1$, $[\sigma^3_y] = \mathbf{g}_2$. Assuming again our Product Rule for vector variables we get: $[\sigma^1_x \sigma^2_y \sigma^3_y] = \mathbf{e}_1 \mathbf{f}_2 \mathbf{g}_2$, and so on (see the first two columns of (38) below). So far, we have not assumed any connection between the orientations formed by the $\mathbf{e}$'s, $\mathbf{f}$'s and $\mathbf{g}$'s. Now, we identify some of the $\mathbf{e}$'s, $\mathbf{f}$'s and $\mathbf{g}$'s, as we did above in the general discussion of (32). (Commutativity of vectors from the relevant orientations is thereby excluded.) Arbitrarily, we choose $i = 1$, $j = 2$ and demand that $\mathbf{e}_1 = -\mathbf{f}_1 = \mathbf{g}_1$ and $\mathbf{e}_2 = \mathbf{f}_2 = \mathbf{g}_2$. In this case, we get $\mathbf{e}_1 \mathbf{f}_2 \mathbf{g}_2 = \mathbf{e}_1 \mathbf{e}_2 \mathbf{e}_2$, and so on (the second and third column of (38) below). Finally, by using (9) and (10), we get the identities forming the third and last column of (38). All in all, we have:

$$[\sigma^1_x \sigma^2_y \sigma^3_y] = \mathbf{e}_1 \mathbf{f}_2 \mathbf{g}_2 = \mathbf{e}_1 \mathbf{e}_2 \mathbf{e}_2 = \mathbf{e}_1 \quad (38a)$$

$$[\sigma^1_y \sigma^2_x \sigma^3_y] = \mathbf{e}_2 \mathbf{f}_1 \mathbf{g}_2 = \mathbf{e}_2 (-\mathbf{e}_1) \mathbf{e}_2 = \mathbf{e}_1 \quad (38b)$$

$$[\sigma^1_y \sigma^2_y \sigma^3_x] = \mathbf{e}_2 \mathbf{f}_2 \mathbf{g}_1 = \mathbf{e}_2 \mathbf{e}_2 \mathbf{e}_1 = \mathbf{e}_1 \quad (38c)$$

$$[\sigma^1_x \sigma^2_x \sigma^3_x] = \mathbf{e}_1 \mathbf{f}_1 \mathbf{g}_1 = \mathbf{e}_1 (-\mathbf{e}_1) \mathbf{e}_1 = -\mathbf{e}_1. \quad (38d)$$

Here our choice of identities among the $\mathbf{e}$'s, $\mathbf{f}$'s and $\mathbf{g}$'s is the counterpart of specifying a certain QM state. As quoted above, Mermin chooses a state $\Phi$ with the property $[\sigma^1_x \sigma^2_y \sigma^3_y] = [\sigma^1_y \sigma^2_x \sigma^3_y] = [\sigma^1_y \sigma^2_y \sigma^3_x] = 1$, which choice implies the property $[\sigma^1_x \sigma^2_x \sigma^3_x] = -1$. We, on the other hand, have assumed identities such that $[\sigma^1_x \sigma^2_y \sigma^3_y] = [\sigma^1_y \sigma^2_x \sigma^3_y] = [\sigma^1_y \sigma^2_y \sigma^3_x] = \mathbf{e}_1$, where $[\sigma^1_x] = -[\sigma^2_x] = [\sigma^3_x] = \mathbf{e}_1$, which equations imply, via the Product Rule, the property $[\sigma^1_x \sigma^2_x \sigma^3_x] = -\mathbf{e}_1$. This parallelism confirms the earlier claim that (38) is a consistent replacement of Mermin's inconsistent (7). Note also that (38) has a simple geometric interpretation: $\Phi$ is a state where the three systems in question all have the same (clockwise or counter-clockwise) orientation in the 1,2-plane (see Appendix B).

Our construction, again, does not explicitly use a locality assumption but it respects (Mermin's version of) such an assumption. Given that we have measured any two of the components of any multivector variable in the first column of (38), we can with certainty predict the third without measuring it. E.g., assume for (38a) that we have found $[\sigma^1_x] = \mathbf{e}_1$ and $[\sigma^2_y] = \mathbf{f}_2 = \mathbf{e}_2$. Then, because the product of all three vectors must be $\mathbf{e}_1$, we can predict that $[\sigma^3_y] = \mathbf{g}_2 = \mathbf{e}_2$ without any measurement of the third system, and similarly for all other components. With Mermin we assume that the three systems are 'mutually well separated' such that there is no influence from the two measured systems onto the unmeasured one. Given this locality assumption we are, as Mermin writes, 'impelled to conclude that the results of measuring either component of any of the three particles must have already been specified prior to any of the measurements' (which is Mermin's formulation of the faithful measurement assumption). Thus, the chosen vector variables must possess all their values jointly, where we have, by our choice of identities between the $\mathbf{e}$'s, $\mathbf{f}$'s and $\mathbf{g}$'s, constraints on these values as specified in the second and third column of (36). In particular, the first three variables (listed in the first column of (38a-c)) must each have the value $\mathbf{e}_1$, while the last must have the value $-\mathbf{e}_1$. But satisfying these constraints is no problem. The equations forming the third and fourth column of (38) are trivial consequences of (9) and (10) and each of these columns multiplies to $-1$.

## VI. DISCUSSION

We have seen that for the PM and GHZ systems a non-contextual or non-contextual and local model can be provided – by exploiting the multivector structure of $\mathbf{G}^3$. This suggestion has a quasi-precedent in a similar approach to Bell's Theorem [31, 32] that is widely criticized as mathematically flawed [33, 34, 35]. This circumstance justifies skepticism against the use of GA concepts in the foundations of QM and requires clarification. To present such clarification, it should be emphasized that in the present approach orientations are interpreted as properties of systems, not of the space accommodating the systems. Moreover, the values of elementary observables (spin components) are interpreted specifically as vectors, not other multivectors (scalars, bivectors or trivectors). Whether this latter assumption is physically meaningful is a separate question (for a brief discussion see Appendix C).

We have not considered Bell's original inequality here. To investigate whether our model can be adapted to the context of Bell's Theorem also is a question to be considered separately. A brief consideration of a special case (the Bell-CHSH inequality in the special case of coplanar vectors) suggests that the present approach can be expanded in this direction (see Appendix D).

## VII. CONCLUSION

We have considered the geometric algebra $\mathbf{G}^3$, which is generated from the vector space $\mathbf{R}^3$. An immediate consequence of the definition of $\mathbf{G}^3$ is equation (8), the fundamental equation of GA for orthonormal vectors. We have interpreted equations concerning unit bi- and trivectors that follow from (8) in terms of 2D and 3D orientations pertaining to different systems. Moreover, we have assumed that elements of two or three different orientations mutually commute iff no identities between any elements of these orientations obtain. From these assumptions plus commutativity, we derived the equations characterizing the PM and GHZ arguments ((4a-f) and (6a-e)), respecting non-contextuality. From these assumptions plus certain identities between elements of different orientations, we derived a proxy for the equations characterizing the Bell-GHZ argument ((7a-d)), respecting locality. Non-contextuality and locality are consequences of the idea of faithful measurement and by salvaging the former we can rebut the arguments against the latter.

Contextuality, non-locality and non-faithful measurement are generally embraced as the key non-classical features of QM. Interpretations trying to restore classicality usually pay a high price in plausibility, as they make extravagant metaphysical assumptions. We have seen, however, that classicality can be recovered at lesser costs, by invoking simple mathematics rather than implausible metaphysics. Moreover, our use of GA implies that the mathematical structure of QM can be further elucidated in terms of geometry. Of course, only after the interpretation thus suggested has been spelled out in more detail, we will be able to judge its merits for our understanding of the theory as a whole.

## APPENDIX A: IDENTITIES OF ORIENTATIONS

In this Appendix we discuss the identity of orientations. As befits a discussion of identity, we assume that the entities considered (here: orientations) are named by individual constants that may or may not refer to the same entity; this assumption makes it meaningful to discuss whether or not two orientations are identical. Assume now that we have two orthonormal bases of $\mathbf{R}^3$, $\{\mathbf{e}_1, \mathbf{e}_2, \mathbf{e}_3\}$ and $\{\mathbf{f}_1, \mathbf{f}_2, \mathbf{f}_3\}$, henceforth briefly the $\mathbf{e}$'s and $\mathbf{f}$'s. We consider orientations constituted solely by $\mathbf{e}$'s or solely by $\mathbf{f}$'s, but no mixed cases. We consider the case where, with respect to an identity of orientations, both orientations are constituted by $\mathbf{e}$'s and the case where one orientation is constituted by $\mathbf{e}$'s, the other by $\mathbf{f}$'s. Since we have 2D and 3D orientations, we get four different cases A1-A4.

### A1. Identity of 2D orientations constituted by e's

Such an identity is claimed in (10), i.e. in $\mathbf{e}_i \mathbf{e}_j = - \mathbf{e}_j \mathbf{e}_i$, for $i \neq j$. In the main text, it is assumed that $\mathbf{e}_i \mathbf{e}_j$ is one of two possible orientations (counter-clockwise and clockwise) in the i,j-plane. This interpretation of the bivector is not standard in the GA literature, where generally $\mathbf{e}_i \mathbf{e}_j$ is assumed to be not an orientation but an oriented area (or plane-segment). The standard interpretation is unfortunate because it does not allow us to understand (10) as a strict identity. We can see this by considering two different methods for generating an area from an ordered pair $(\mathbf{a}, \mathbf{b})$ of linearly independent vectors $\mathbf{a}, \mathbf{b}$. First, we can assume such an area to be a circle sector that is generated by rotating, say, $\mathbf{a}$ until it coincides with $\mathbf{b}$, where the area covered by $\mathbf{a}$ during the translation is the bivector $\mathbf{a} \mathbf{b}$. By this method, however, $\mathbf{e}_i \mathbf{e}_j$ and $(- \mathbf{e}_j) \mathbf{e}_i$ are not identical. Assuming that for $\mathbf{e}_i \mathbf{e}_j$ with $i = 1, j = 2$, $\mathbf{e}_1$ points, say, to 3 o'clock on the unit circle and $\mathbf{e}_2$ to 12 o'clock. The covering motion leading to $\mathbf{e}_1 \mathbf{e}_2$ is counter-clockwise and $\mathbf{e}_1 \mathbf{e}_2$ is the unit circle quadrant between 3 and 12 o'clock. By the same method, $(- \mathbf{e}_2) \mathbf{e}_1$ is the area between 6 and 3 o'clock, which evidently is a different quadrant of the unit circle. The second method to generate an area from $(\mathbf{a}, \mathbf{b})$ is to form a parallelogram from them by translating, say, $\mathbf{a}$ from the base of $\mathbf{b}$ to its tip, where the area covered by $\mathbf{a}$ is the bivector $\mathbf{a} \mathbf{b}$. By this method, $(\mathbf{e}_1, \mathbf{e}_2)$ generates a square in the 3 o'clock-12 o'clock quadrant of the unit circle, with the origin in its lower left corner, but $(- \mathbf{e}_2, \mathbf{e}_1)$ generates a square in the 6 o'clock-3 o'clock quadrant, with the origin in its upper left corner; again, the two areas are not identical, which forbids calling them so.

In view of this difficulty, we avoid the usual interpretation of the bivector $e_i\, e_j$ as an oriented area in favor of its interpretation as an orientation. For preparation, we recall the traditional geometric definition of a vector as a certain class. Thus, a Euclidean vector is defined as an equivalence class, under equipollence (having the same length and direction), of directed line segments in a Euclidean space (like $\mathbf{R}^3$). We call such a class a *free vector* (written as e.g. '<**a**>') and call any element of such a class, i.e. any line segment with a certain fixed length and direction, a bound vector (written as e.g. '**a**'). Thus, vector **a**, but not vector <**a**>, may be identified with an ordered pair of points; vector **a** is localized, but vector <**a**> is not.

In order to define the orientation of an area, we must address how it has been generated. Let, for two linearly independent vectors **a** and **b** that lie in the i,j-plane and are bound to a point P identical with both their bases, the **a**,**b**-area with respect to P be the area covered by translating **a** from the base of **b** to its tip, such that after the translation the base of **a** is no longer identical with P, while the one of **b** still is. Let the tips of **a** and **b** before a translation be points A and B and the tips of **a** and **b** after a translation be points A′ and B′. Then the **a**,**b**-area and the **b**,**a**-area are the parallelograms P-A-A′-B and P-A-B′-B, respectively (where the points defining the parallelograms are mentioned counter-clockwise). Since A′ = B′, these parallelograms are the same figure. But this figure has been generated by translating **a** along **b**, or, alternatively, by translating **b** along **a**. We thus define for two vectors **a** and **b** bound to point P as their common base, the **a**,**b**-area with respect to point P as the area covered by translating **a** along **b**, where the understanding is that not only would the area be generated by that translation but that it has been so generated. Then, assuming that the **a**,**b**-area has been generated by translating **a** along **b**, we have the desired result that the **a**,**b**-area and the **b**,**a**-area with respect to the same point are not identical.

For a full definition of a 2D orientation, we require further ancillary definitions. Let an arbitrary **a**,**b**-area be the **a**,**b**-area with respect to an arbitrary point. (Given an arbitrary point P and a free vector <**a**>, the latter contains exactly one bound vector **a** ∈ <**a**> with P as its base, and similarly for **b**.) Let the primitive **a**,**b**-area be the **a**,**b**-area with respect to the origin O. Let, for two linearly independent vectors **c** and **d**, an **a**,**b**-area be an arbitrary **c**,**d**-area that can be identified with the primitive **a**,**b**-area by means of a translation and a rotation around O such that the point sequences (O, C, C′) and (O, A, A′) coincide, i.e. C = A, C′ = A′. Finally, define an orientation in the i,j-plane to be the class of $e_i,e_j$-areas. Hence, when interpreting the bivector $e_i\, e_j$ not as an area but as an orientation we interpret it as a certain class of areas. This interpretation now allows us to identify the bivectors in (10): $e_i\, e_j$ and $-\, e_j\, e_i$ are identical in the sense that they are the same class of areas.

This construction also allows an understanding of how vectors *constitute* bivectors. When we say that $e_i$, $e_j$ plus the left-right ordering for their product, constitute the bivector $e_i\, e_j$, this abbreviates a more involved structure. The vectors $e_i$, $e_j$, if the bivector $e_i\, e_j$ is to be constructed from them, are vectors bound to O, an $e_i,e_j$-area is an arbitrary $e_i,e_j$-area identifiable, by translation and rotation, with the primitive $e_i,e_j$-area. The bivector $e_i\, e_j$, the orientation, is the class of $e_i,e_j$- areas, and is *constituted* by these areas. Since these areas are identical with the $-\, e_j\, e_i$-areas, so are the constituted classes. We can briefly speak of vectors constituting bivectors and can, e.g., say that the vectors $e_i$, $e_j$, plus the left-right ordering for their product, constitute the bivector $e_i\, e_j$, and so do the vectors $-\, e_j$, $e_i$, plus the same ordering.

Thus: the vectors $e_i$, $e_j$, plus the left-right ordering for their product, constitute the bivector $e_i\, e_j$, and so do the vectors $-\, e_j$, $e_i$, plus the same ordering. More generally, the four bivectors $e_i\, e_j$, $(-\, e_j)\, e_i$, $(-\, e_i)\, (-\, e_j)$ and $e_j\, (-\, e_i)$ are the same bivector, thus the same orientation. So, the four bivectors are constituted by different vectors (plus the left-right ordering) and yet are the same bivector, as claimed in the main text. It is easy to see that, all in all, using the four vectors $e_i$, $e_j$, $-\, e_i$, $-\, e_j$, we get eight possibly different orientations that, due to identity claims as in (10), turn out to be just two different ones: the *counter-clockwise* and *clockwise* orientation.

**A2. Identity of 3D orientations constituted by e's**

Such an identity is claimed in $e_i\, e_j\, e_k = -\, e_j\, e_i\, e_k$, which follows from the trivial (13) ($e_i\, e_j\, e_k = e_i\, e_j\, e_k$) above, by (10). We proceed as in Appendix A1. Let, for three linearly independent vectors **a**, **b**, **c** in $\mathbf{R}^3$ that are bound to a point P identical with their three bases, the **a**,**b**,**c**-volume with respect to P be the volume covered by translating the **a**,**b**-area with respect to P from the base of **c** to its tip, such that after the translation the bases of **a** and **b** are no longer identical with P, while the one of **c** still is. Hence, the **a**,**b**,**c**-volume has been generated by translating the **a**,**b**-plane along **c**, instead of, alternatively, by translating another plane (e.g., the **a**,**c**-plane) along a vector not coplanar with it (e.g. **b**). We thus define the **a**,**b**,**c**-area with respect to P as the area covered by translating the **a**,**b**-plane along **c**, with the interpretation that the volume has indeed be so generated. Now, as before, let an arbitrary **a**,**b**,**c**-volume be the **a**,**b**,**c**-volume with respect to an arbitrary point. Let the primitive **a**,**b**,**c**-volume be the **a**,**b**,**c**-volume with respect to the origin O. Let, for three vectors **d**, **e**, **f**, an **a**,**b**,**c**-volume be an arbitrary **d**,**e**,**f**-volume that can be identified with the primitive **a**,**b**,**c**-volume by means of a translation and a rotation around O such that the point sequences (O, D, D′, D″) and (O, A, A′, A″) coincide, i.e. D = A, D′ = A′, D″ = A″. (Here, A″ is the tip of the vector resulting from translating **a** along **b** and then the **a**,**b**-plane along **c**. Similarly, for **d**, **e**, **f** and D″.) Finally, define an orientation in $\mathbf{R}^3$ to be the class of $\mathbf{e}_i,\mathbf{e}_j,\mathbf{e}_k$-volumes. Hence, when interpreting the trivector $\mathbf{e}_i\,\mathbf{e}_j\,\mathbf{e}_k$ not as an oriented volume but as an orientation we interpret it as a certain class of volumes. This allows us to understand $\mathbf{e}_i\,\mathbf{e}_j\,\mathbf{e}_k = -\,\mathbf{e}_j\,\mathbf{e}_i\,\mathbf{e}_k$ strictly as an identity. All identities derivable from (13) jointly show that there are only two different orientations, the *right-handed* and the *left-handed* one.

### A3. Identity of 2D orientations constituted by e's and by f's

Such an identity is considered in the discussion after (23) above, where it is claimed that this identity leads to inconsistency when we assume the **e**'s and **f**'s to commute. Here is the proof. Assume $\mathbf{e}_i\,\mathbf{e}_j = \mathbf{f}_i\,\mathbf{f}_j$ and commutativity of **e**'s and **f**'s. From $\mathbf{e}_i\,\mathbf{e}_j = \mathbf{f}_i\,\mathbf{f}_j$, on the one hand, by left-multiplication of $\mathbf{e}_j$, we get $-\mathbf{e}_i = \mathbf{e}_j\,\mathbf{f}_i\,\mathbf{f}_j$; on the other hand, by right-multiplication of $\mathbf{e}_j$, we get $\mathbf{e}_i = \mathbf{f}_i\,\mathbf{f}_j\,\mathbf{e}_j$ and, by commutativity, $\mathbf{e}_i = \mathbf{e}_j\,\mathbf{f}_i\,\mathbf{f}_j$. Hence, $\mathbf{e}_i = \mathbf{0}$, which contradicts the initial assumption that $\mathbf{e}_i$ is a unit vector. Thus, $\mathbf{e}_i\,\mathbf{e}_j = \mathbf{f}_i\,\mathbf{f}_j$ and commutativity of **e**'s and **f**'s are incompatible.

### A4. Identity of 3D orientations constituted by e's and by f's

Such an identity is claimed in (20) above: $\mathbf{e}_i\,\mathbf{e}_j\,\mathbf{e}_k = \mathbf{f}_l\,\mathbf{f}_m\,\mathbf{f}_n$, where the triple (l, m, n) is any permutation of (1, 2, 3). Assume that (l, m, n) is an even permutation of (1, 2, 3). Then, by using (10), we can show that $\mathbf{f}_l\,\mathbf{f}_m\,\mathbf{f}_n = \mathbf{f}_i\,\mathbf{f}_j\,\mathbf{f}_k$. Assume alternatively that (l, m, n) is an odd permutation of (1, 2, 3). Then, again by using (10), we can show that $\mathbf{f}_m\,\mathbf{f}_l\,\mathbf{f}_n = \mathbf{f}_i\,\mathbf{f}_j\,\mathbf{f}_k$. We can thus dispense with the indices (l, m, n). Let (20) be $\mathbf{e}_i\,\mathbf{e}_j\,\mathbf{e}_k = \mathbf{f}_i\,\mathbf{f}_j\,\mathbf{f}_k$. Then $\mathbf{f}_i\,\mathbf{f}_j\,\mathbf{f}_k$ is the same handedness as $\mathbf{e}_i\,\mathbf{e}_j\,\mathbf{e}_k$. Let (20) be $\mathbf{e}_i\,\mathbf{e}_j\,\mathbf{e}_k = \mathbf{f}_j\,\mathbf{f}_i\,\mathbf{f}_k$. Then $\mathbf{f}_j\,\mathbf{f}_i\,\mathbf{f}_k$ is the same handedness as $\mathbf{e}_i\,\mathbf{e}_j\,\mathbf{e}_k$. Without loss of generality, we may choose between both cases and arbitrarily choose the first such that (20) simplifies to (21): $\mathbf{e}_i\,\mathbf{e}_j\,\mathbf{e}_k = \mathbf{f}_i\,\mathbf{f}_j\,\mathbf{f}_k$.

What does it mean to say that $\mathbf{e}_i\,\mathbf{e}_j\,\mathbf{e}_k = \mathbf{f}_i\,\mathbf{f}_j\,\mathbf{f}_k$? In the main text, we interpret unit bi- and trivectors not as orientations characterizing a certain plane in $\mathbf{R}^3$ or the whole of $\mathbf{R}^3$, but as orientations characterizing *systems* extended in a plane or in $\mathbf{R}^3$. Hence, we can distinguish two cases of the identity of orientations: (a) a case where two orientations of the same system are identical and (b) a case where two orientations of different systems are identical. In other words, the **e**'s and **f**'s in an identity claim may belong to the same or different systems.

Before we consider this distinction abstractly, we note that it is intuitively accessible. Consider a single glove. We can meaningfully ask whether orientations given by the sequences (thumb, forefinger, middle finger) and (middle finger, thumb, forefinger) are identical. (Intuitively, they are.) On the other hand, consider a pair of gloves. Here we can ask whether orientations given by the sequences (left thumb, left forefinger, left middle finger) and (right thumb, right forefinger, right middle finger) are identical. (Intuitively, they are not.) It is this distinction of identities of orientations in the same system or different systems that we must consider now.

Let an orientation constituted only by **e**'s and an orientation constituted only by **f**'s both pertain to the same system. Assume (as is done in the main text) that a system has a unique orientation. Then the **e**- and the **f**-orientation are identical. We can thus drop one of the bases and stipulate that the **e**- and the **f**-orientation always pertain to different systems (as is done in the main text).

Consider now two orientations – the **e**- and the **f**-orientation – pertaining to different systems S1 and S2. We may assume that these systems are compact subsets of $\mathbf{R}^3$, that they are disjoint and that their respective orientations are associated with them, individually. Nonetheless, since these orientations are classes of certain volumes, two orientations can be strictly identical and that they are so is the claim of (21). We may consider two points P and Q (neither of which has to be identical with O) within S1 and S2, respectively, and consider the $\mathbf{e}_i\,\mathbf{e}_j\,\mathbf{e}_k$-volume with respect to P and the $\mathbf{f}_i\,\mathbf{f}_j\,\mathbf{f}_k$-volume with respect to Q; since the **e**'s and **f**'s making up the volumes are bound, respectively, to P and Q, these two volumes are non-identical, in contrast with the **e**- and **f**-orientation that are identical.

Consider finally the case where the **e**- and the **f**-orientation are identical and certain identities among their component vectors hold. The vectors can be identical iff they are free vectors, independently of the fact that the **e**- and **f**-orientation they pertain to were constructed from bound vectors that are not identical. We may interpret an identity $\mathbf{e}_i = \mathbf{f}_i$ mentioned directly above (32) in two ways, one being the description as an identity of classes (of bound vectors): $\langle\mathbf{e}_i\rangle = \langle\mathbf{f}_i\rangle$. Alternatively, consider that, given $\langle\mathbf{e}_i\rangle = \langle\mathbf{f}_i\rangle$, for every point P, every vector $\mathbf{e}_i$ bound to P and every vector $\mathbf{f}_i$ bound to Q, $\langle\mathbf{f}_i\rangle$ contains exactly one vector $\mathbf{f}_i{'} \in \langle\mathbf{f}_i\rangle$ having P as its base and satisfying the identity $\mathbf{e}_i = \mathbf{f}_i{'}$. So, for every vector $\mathbf{e}_i$ bound to P there is an element of $\langle\mathbf{f}_i\rangle$, though one bound to P, not Q, that is identical with it.

### APPENDIX B: THE QM STATES Φ AND Φ′

We want to give a geometric interpretation of (38) and (32), where the QM eigenstates of the chosen observables and values are $\Phi = 1/\sqrt{2}\,(|+\;+\;+\rangle - |-\;-\;-\rangle)$ and $\Phi' = 1/\sqrt{2}\,(|+\;-\;+\rangle + |-\;+\;-\rangle)$, respectively. First, consider that the rightmost column of (38a-d) is the sequence $(\mathbf{e}_1, \mathbf{e}_1, \mathbf{e}_1, -\mathbf{e}_1)$ iff the identities chosen in the main text, i.e.

$e_1 = (-f_1) = g_1$ and $e_2 = f_2 = g_2$, hold. These identities are sufficient, but not necessary, to specify a case where the rightmost column of (38a-d) multiplies to $(-1)$. E.g., choosing identities $e_1 = f_2 = g_1$ and $e_2 = f_1 = g_2$) we find that column to be $g_2, g_2, g_2, (-g_2)$, which again multiplies to $-1$. Indeed, for any vector $x$ with $x \in \{\pm e_1, \pm e_2, \pm f_1, \pm f_2, \pm g_1, \pm g_2\}$ we can find identities such that the last column of (38) takes the form $(x, x, x, -x)$ and thus multiplies to $-1$. This shows that our choice of identities leading, in (38), to the sequence $(e_1, e_1, e_1, -e_1)$ as replacement for Mermin's $(1, 1, 1, -1)$ on the right of (7) was truly arbitrary.

We have seen that, while (7) cannot be satisfied, a variant of (7), wherein real numbers are replaced by vectors, can. However, this is done, in (38a-d), in a manner that seems entirely ad hoc. In particular, every choice of a vector $x$ and identities implying a sequence $(x, x, x, -x)$ for the last column of (38a-d), seems to entail orientations in the plane that have an implausible structure. E.g., the first set of identities above, i.e. $e_1 = -f_1 = g_1$ and $e_2 = f_2 = g_2$, seems to directly imply that we have orientations $e_1 e_2, (-f_1) f_2, g_1 g_2$ such that $e_1 e_2$ and $g_1 g_2$ are the same (e.g., the counter-clockwise) orientation, while $(-f_1) f_2$, is a different (given the previous choice: the clockwise) orientation. If this really were the case it would be a strange result, for Mermin's $\Phi = 1/\sqrt{2} |+++> - |--->$) evidently is a QM state that mathematically represents all three spins alike such that any hidden-variables theory representing them differently is implausible.

But this complaint ignores the nature of our hidden variables: orientations. Assume that we first have chosen $x = e_1$ and thus intend to find identities such that the last column of (38) is the sequence $(e_1, e_1, e_1, -e_1)$. We work out the identities and find $e_1 = -f_1 = g_1$ and $e_2 = f_2 = g_2$. The identities $f_2 = e_2$ and $f_1 = -e_1$ then yield $e_1 e_2 = f_2 f_1$, which says that $e_1 e_2$ and $f_2 f_1$ are identical, i.e. are the same orientation. But what guarantees that $f_2 f_1$ is the orientation *of the second system*? To answer this question, we adopt the following criterion. A bivector $e_i e_j$ is a system's orientation in the i,j-plane iff the vector variables $\sigma_i, \sigma_j$ are determined in the order $(\sigma_i, \sigma_j)$ with values $[\sigma_i] = e_i$ and $[\sigma_j] = e_j$. (A vector variable is determined iff it is given a value.) We assume that in (38) the components $\sigma^1_x, \sigma^2_y, \sigma^3_y$ of $\sigma^1_x \sigma^2_y \sigma^3_y$ are determined jointly, i.e. $[\sigma^1_x] = \pm e_1$ iff $[\sigma^2_y] = \pm f_2$, and analogously for the other components and multivectors. Now assume that $e_1 e_2$ is the first system's orientation. This implies that $\sigma^1_x, \sigma^1_y$ have been determined in the order $(\sigma^1_x, \sigma^1_y)$ (i.e. top-down in (38a-d)) with values $[\sigma^1_x] = e_1$ and $[\sigma^1_y] = e_2$. The first of these determinations entails that also $\sigma^2_y$ has been determined, i.e. $[\sigma^2_y] = \pm f_2$. In (38), we have assumed $[\sigma^2_y] = f_2$. Similarly, the determination of $\sigma^1_y$ implies $[\sigma^2_x] = f_1$. Thus, the variables $\sigma^2_y$ and $\sigma^2_x$ are determined in the order $(\sigma^2_y, \sigma^2_x)$ (top-down in (38a-d)) with the results $f_2$ and $f_1$, in this order. (The order of determination has been transferred from the first to the second system.) Hence, $f_2 f_1$ is the second system's orientation. Consider now the identities $f_2 = e_2$ and $f_1 = -e_1$. They imply that $f_2 f_1 = e_2 (-e_1) = e_1 e_2$, which is to say that the first and second system's orientations, $e_1 e_2$ and $f_2 f_1$, are identical.

For contrast, consider $[\sigma^3_x]$ and $[\sigma^3_y]$. The order in $e_1 e_2$ and $f_2 f_1$ or, more generally, the top-down order in (38) implies nothing about their values' order and thus nothing about the orientations $g_1 g_2$ and $g_2 g_1$. But our task of finding identities such that rightmost column in (38) is given by the sequence $(e_1, e_1, e_1, -e_1)$ has yielded that either $g_2 = e_2$ and $g_1 = e_1$ or $g_2 = -e_2$ and $g_1 = -e_1$. This suffices to infer $e_1 e_2 = g_1 g_2$, which implies that the first and third system's orientation are identical.

As a result of these considerations, the three systems' orientations are $e_1 e_2, f_2 f_1, g_1 g_2$, where $e_1 e_2 = f_2 f_1 = g_1 g_2$, such that all three systems have the same orientation in the 1,2-plane. Hence, we can give a simple geometric interpretation of the QM state $\Phi = 1/\sqrt{2} |+++> - |--->$) reflecting its internal symmetry: $\Phi$ is a state where the three systems have the same orientation in the 1,2-plane, either clockwise or counter-clockwise.

To corroborate this interpretation, consider the QM state associated with (32a-d). We begin with Mermin's four operators $\sigma^1_x \sigma^2_y \sigma^3_y, \sigma^1_y \sigma^2_x \sigma^3_y, \sigma^1_y \sigma^2_y \sigma^3_x, \sigma^1_x \sigma^2_x \sigma^3_x$, which have a common eigenstate $\Phi'$ such that the eigenvalues are 1, $(-1)$, 1, 1, respectively. We find $\Phi' = 1/\sqrt{2} (|+-+> + |-+->)$, a state that, in contrast with $\Phi$, represents different particles differently. The geometric interpretation of $\Phi'$ reflects this feature. Now, to satisfy the rightmost column of (32), we require identities such that $e_1 f_2 g_2 = -(e_2 f_1 g_2) = e_2 f_2 g_1 = e_1 f_1 g_1 = e_1$ and find $e_1 = f_1 = g_1$ and $e_2 = f_2 = g_2$. (Note that above, in the paragraph before (32), we proceeded in the reverse order, assumed the identities and derived the sequence $(e_1, -e_1, e_1, e_1)$, an instance of the rightmost column of (32).)

To find the systems' actual orientations, we again apply our criterion. Assume again that $e_1 e_2$ is the first system's orientation. This implies that $\sigma^1_x, \sigma^1_y$ have been determined in the order $(\sigma^1_x, \sigma^1_y)$ (top-down in (32a-d)) with values $[\sigma^1_x] = e_1$ and $[\sigma^1_y] = e_2$. The first of these determinations entails that also $\sigma^2_y$ has been determined, i.e. $[\sigma^2_y] = \pm f_2$. In (32), we have effectively assumed $[\sigma^2_y] = f_2$. Similarly, the determination of $\sigma^1_y$ implies $[\sigma^2_x] = f_1$. Thus, $\sigma^2_y$ and $\sigma^2_x$ are determined in the order $(\sigma^2_y, \sigma^2_x)$ (top-down in (32a-d)) with the results $f_2$ and $f_1$, such that $f_2 f_1$ is the second system's orientation. Now, however, we have the identities $f_2 = e_2$ and $f_1 = e_1$. They imply that $f_2 f_1 = e_2 e_1$. This orientation is non-identical with the assumed orientation of the first system and thus we have the result that the first and second system's orientations, $e_1 e_2$ and $f_2 f_1$, are non-identical.

The considerations about [$\sigma^3_x$] and [$\sigma^3_y$] are the same as before and again yield that the third system's orientations are $g_1 g_2$, with $e_1 e_2 = g_1 g_2$, which again implies that the first and third system's orientations are identical. So, these latter orientations are identical and differ from the second system's orientation. E.g. the first and third systems might have a counter-clockwise orientation, in which case the second system's orientation is clockwise. As announced, we see a geometric illustration of how the QM state $\Phi' = 1/\sqrt{2}$ (|+ − +> + |− + −>) represents different systems differently.

## APPENDIX C: VALUES OF OBSERVABLES AS VECTORS

The interpretation of values of observables as vectors is based on the assumption that every QM system is equipped with an orientation such that its components are the observables' values. But this latter assumption is itself a mere assumption. Thus, the question arises whether that interpretation is plausible in itself: *should* the values of QM observables be interpreted as vectors? Consider the observable 'spin in the x-direction', where x is any vector in $\mathbf{R}^3$. Since x is directed, it seems that the observable is directed such that its value cannot have an extra direction on top and so the latter must simply be a scalar. Now, it is not the observable alone but the observable's value that exists in physical reality, i.e. in the physical system. If we consider the value to be a scalar, we lose the distinction between spin components and observables whose values really are scalars (like energy), so we should take the spin component to be a vector. But then should not the observable 'spin in the x-direction' just be a dimension (the space spanned by x) and thus not an entity in whose name the word 'direction' appears? Consider a dimension and two distinguishable points that we can put into it and that we name +1 and − 1. Both from an internal as well as external standpoint, we have two options to arrange them (first: 1, second: − 1, or vice versa; left: 1, right: − 1 or vice versa). We can choose one option and by that choice have created a directed dimension, an entity adequately represented by a vector, although its value is not a scalar but a vector again. We may name the spin observables' value + 1 and − 1, but should think of these scalars as representing the values 'parallel' and 'antiparallel' of an entity that has direction: a vector variable or observable. A vector and a scalar cannot be parallel or antiparallel but two vectors can.

## APPENDIX D: RELATION TO THE BELL-CHSH-INEQUALITY

We consider the Bell-CHSH-inequality and its derivation as presented by Redhead [36]. We consider a two-particle spin-½ system with $\mathbf{a}$, $\mathbf{a'}$ and $\mathbf{b}$, $\mathbf{b'}$ being directions pertaining, respectively, to systems $\mathbf{A}$ and $\mathbf{B}$. Let $a_n$ be the value (parallel or antiparallel) of the spin-component $\mathbf{a}$ of $\mathbf{A}$, which is possessed by $\mathbf{A}$ and faithfully revealed in the n-th measurement of the value of $\mathbf{a}$; similarly for $a'_n$, $b_n$, $b'_n$. From $a_n$, $a'_n$, $b_n$, $b'_n$ form the expression:

$$\gamma_n = a_n b_n + a_n b'_n + a'_n b_n - a'_n b'_n. \quad (C1)$$

Thus:

$$\gamma_n = a_n (b_n + b'_n) + a'_n (b_n - b'_n). \quad (C2)$$

Assume that $a_n$, $b_n$, $a'_n$, $b'_n \in \{1, -1\}$. Then from (C2):

$$\gamma_n = \pm 2. \quad (C3)$$

Now consider N events and form the expression:

$$|\frac{1}{N}\sum_{n=1}^{N}\gamma_n| = |\frac{1}{N}\sum_{n=1}^{N} a_n b_n + \frac{1}{N}\sum_{n=1}^{N} a_n b'_n$$
$$+ \frac{1}{N}\sum_{n=1}^{N} a'_n b_n - \frac{1}{N}\sum_{n=1}^{N} a'_n b'_n|. \quad (C4)$$

From (C3), we have $|\frac{1}{N}\sum_{n=1}^{N}\gamma_n| \leq 2$. Define the classical correlation $c(\mathbf{a}, \mathbf{b}) = \lim_{N\to\infty} \frac{1}{N}\sum_{n=1}^{N} a_n b_n$ and similarly for $a_n b'_n$, $a'_n b_n$, $a'_n b'_n$. Then for $N \to \infty$, we have the CHSH inequality:

$$|c(\mathbf{a}, \mathbf{b}) + c(\mathbf{a}, \mathbf{b'}) + c(\mathbf{a'}, \mathbf{b}) - c(\mathbf{a'}, \mathbf{b'})| \leq 2. \quad (C5)$$

Consider now the QM correlations $c_{qm}(\mathbf{a}, \mathbf{b})$, etc. As is well-known, for the singlet state one finds $c_{qm}(\mathbf{a}, \mathbf{b}) = -\cos\theta_{ab}$, etc. For simplification, let $\mathbf{a}$, $\mathbf{b}$, $\mathbf{a'}$, $\mathbf{b'}$ be coplanar with $\theta_{ab} = 0$, $\theta_{ab'} = \theta_{a'b} = \varphi$, $\theta_{a'b'} = 2\varphi$ (as in [36]). Then, because of the $c_{qm}(\mathbf{a}, \mathbf{b})$, etc., the LHS of (C5) must be a function F: $[0, \pi]$ of $\varphi$ of the form:

$$F(\varphi) := |1 + 2\cos\varphi - \cos 2\varphi| \quad (C6)$$

If the classical and QM correlations are identical, we must have $F(\varphi) \leq 2$ for all $\varphi$. But we have, e.g., $F(\pi/4) = 1 + \sqrt{2}$ and $F(\pi/3) = 5/2$. Since $F(\pi/3)$ is the maximum of $F(\varphi)$, we have $F(\varphi) \leq 5/2$ for all $\varphi$.

We have assumed (in sec. IV) that each subsystem of a QM system is characterized by a certain orientation, such that the orientation's components are vectors equipped with an order-sensitive product (i.e. obeying (8) above). In the present Bell-CHSH case, we simply assume that the QM observables and their values are replaced by vector variables and vector values obeying (8). Thus, assume that in (C1) $a_n$, $b_n$, $a'_n$, $b'_n$ are not scalars but vectors in $\mathbf{R}^3$, where $\mathbf{R}^3$ generates $\mathbf{G}^3$. Then these vectors are also elements of $\mathbf{G}^3$; moreover, $\gamma_n$ and its summands are even multivectors (sums of scalars and pure bivectors) in $\mathbf{G}^3$. Since now $\gamma_n$ is not generally a scalar, (C3) fails. Accordingly, (C5) cannot be derived. In particular, by our choice of $\mathbf{b}$ and $\mathbf{b'}$, $b_n$ and $b'_n$ are non-collinear, thus in (C2) neither $(b_n + b'_n) = 0$ nor $(b_n - b'_n) = 0$.

Moreover, if $a_n$, $b_n$, $a'_n$, $b'_n$ are vectors, then, with the above choices for $\theta_{ab}$, $\theta_{ab'}$, $\theta_{a'b}$, $\theta_{a'b'}$, they can be written as follows: $a_n = b_n = e_1 \cos \varphi + e_3 \sin \varphi$, $a'_n = e_1 \cos 2\varphi + e_3 \sin 2\varphi$, $b'_n = e_1$, such that (C1) becomes:

$$\gamma_n = 1 + 2 \cos \varphi - \cos 2\varphi + (2 \sin \varphi - \sin 2\varphi)\, e_3 e_1. \qquad (C7)$$

Make the plausible assumption that the correlation of two vectors a and b is given by the absolute value of the scalar part of the multivector ab, which value we write as $\|ab\|$. Then, from (C7):

$$\|\gamma_n\| = |1 + 2 \cos \varphi - \cos 2\varphi|. \qquad (C8)$$

Hence, the assumption that observables' values are vectors in $\mathbf{G}^3$ plus the assumption just made concerning correlations allow us to exactly reproduce the QM bound 5/2 for the left side of (C5).

---